\documentclass[aps,pra,showpacs,twocolumn,eqsecnum]{revtex4}
\usepackage{bm, graphics, amsmath}
\begin{document}

\title{Optimized maximum-confidence discrimination of $N$ mixed quantum states and application to symmetric states}

\author{Ulrike Herzog}
\affiliation {Nano-Optics, Institut f\"ur Physik,
Humboldt-Universit\"at Berlin, Newtonstrasse 15, D-12489 Berlin,
Germany}

\begin{abstract}
We study an optimized measurement which discriminates $N$  mixed quantum states occurring with given prior probabilities. 
The measurement yields the maximum achievable confidence for each  of the $N$ conclusive outcomes,
thereby keeping the overall probability of inconclusive outcomes as
small as possible.  It corresponds to 
optimum unambiguous discrimination  when  for each outcome the confidence is equal to unity. Necessary and
sufficient optimality conditions are derived and  general
properties of the optimum measurement are obtained.  The  results are  
applied to the optimized maximum-confidence discrimination of $N$ equiprobable 
symmetric mixed states. Analytical solutions are presented  for 
a number of examples,  including the discrimination of $N$ symmetric pure states 
spanning a $d$-dimensional Hilbert space ($d\leq N$) and  the discrimination 
of $N$ symmetric mixed  qubit states.  
\end{abstract}


\maketitle

\section{Introduction}
The discrimination between different quantum states is an essential task in quantum communication, quantum cryptography and quantum computing. Since nonorthogonal states cannot be distinguished
perfectly, various optimized discrimination strategies have been
developed.
The two  best known of these are minimum-error discrimination
\cite{helstrom} and optimum unambiguous discrimination
 \cite{ivan,dieks,peres,jaeger}. In the latter strategy, errors are not allowed, at the expense of
admitting inconclusive results the probability of which  is minimized
in the optimum measurement. Recently the optimum unambiguous discrimination between mixed states attracted a lot of interest
\cite{rudolph,raynal,eldar,HB,BFH,herzog,raynal2,kleinmann,feng1,zhang}.
 Unambiguous discrimination of each single state in a given set of states is only possible when  pure states are linearly independent \cite{chefles1},
and when for  mixed states the support \cite{support} of each respective density operator is 
different from the support of every other density operator belonging to a state in the  set \cite{rudolph,raynal}. 

For the case when it is not possible to discriminate  each single state  unambiguously,  Croke and co-workers
\cite{croke,croke1} introduced the strategy of maximum-confidence  discrimination which was  experimentally demonstrated  for
 three symmetric pure qubit states \cite{mosley}.
In this strategy,   we are as confident as possible that  the respective state was
indeed present when a conclusive  result is obtained.   Maximum-confidence discrimination corresponds to  unambiguous discrimination when for each conclusive result the  confidence  is equal to unity.
  As with unambiguous discrimination, also with maximum-confidence discrimination in
general the measurement is not unique and a further optimization can be performed which minimizes the 
   number of inconclusive results, or, in other words, the failure probability of the discrimination measurement.  
The optimized  maximum-confidence discrimination of two mixed
qubit states  has been theoretically investigated in our previous paper \cite{herzog1}. 
We also proposed an implementation of the optimum measurement 
 \cite{herzog-benson} which has been recently experimentally
realized for two mixed qubit states having the same purity \cite{steudle}.

In the present paper we extend the theoretical investigations  
 to the discrimination of an arbitrary number of mixed states. 
 In Sec. II we describe the measurement for maximum-confidence discrimination of $N$ mixed
states. Section III is devoted to the problem of
determining the optimum  measurement which yields
the smallest failure probability. Necessary and
sufficient optimality conditions are derived and  general
properties of the optimum measurement are obtained. 
A proof concerning the necessity of the optimality
conditions is given in the Appendix. In Sec.
IV we apply our results to the discrimination of symmetric
states and obtain analytical solutions for special cases. Section V  concludes the paper. 
We add that recently a related publication appeared where 
maximum-confidence discrimination is treated for $N$ symmetric pure 
qudit states \cite{jimenez}.  

\section{Detection operators for maximum-confidence discrimination}

We suppose that a quantum system is  prepared with the prior
probability $\eta_j$ in one of $N$ given states described by the
density operators $\rho_j$ $(j=1,\ldots,N)$, where we assume that
$\sum_{j=1}^N \eta_j =1$. It will be convenient to introduce the
density operator $\rho$ characterizing the total information about
the quantum system,
\begin{equation}
\label{rho-x}
 \rho=  \sum_{j=1}^N \eta_j \rho_j=\sum_{l=1}^d r_l|r_l\rangle \langle
 r_l|\quad{\rm with} \;\;\sum_{l=1}^d |r_l\rangle \langle r_l| = I_d,
\end{equation}
where for later use we introduced the spectral representation of
$\rho$. The eigenstates $|r_l\rangle$ form a complete orthonormal
basis in the $d$-dimensional  Hilbert space
 ${\cal H}_d$ jointly spanned by the eigenstates  of $\rho_1, \ldots \rho_N$ that  belong to nonzero eigenvalues,
and $I_d$ is  the identity operator in ${\cal H}_d$. 
 We want to perform a measurement
in order to infer from a single outcome in which of the $N$ possible
states the system was prepared.
 The discrimination measurement is described by $N+1$ positive detection
 operators $\Pi_0,\Pi_1,\ldots, \Pi_N$  fulfilling the completeness relation
$\sum_{j=0}^N \Pi_j=I_d$. The conditional probability that a system
 is inferred to be in the state $\rho_j$  given it has been prepared
 in the state $\rho_k$ reads
$p(j\,|\rho_k\,) = {\rm Tr}(\rho_k\Pi_j)$,
  while ${\rm Tr}(\rho_k\Pi_0)$
is the conditional probability that the measurement yields an
inconclusive result.
From the completeness
relation we get the requirement
\begin{eqnarray}
\label {cond1} \Pi_0=I_d-\sum_{j=1}^N \Pi_j \geq 0, \quad\Pi_j \geq
0 \quad(j=1,\ldots,N).\quad
\end{eqnarray}

 The confidence in the conclusive
measurement outcome $j$ has been introduced as the conditional
probability $p(\rho_j\,|j\,)= p(\rho_j,j)/{p(j)}$ that the state
$\rho_j$ was indeed prepared, given that the outcome $j$ is
detected  \cite{croke}. Here $p(\rho_j,j)= \eta_j {\rm Tr} (\rho_j\Pi_j)$ is the
joint probability that the state $\rho_j$ was prepared and the
detector $j$ clicks, and
$p(j)= {\rm Tr} (\rho\Pi_j)$
is the total probability for the detection of the outcome $j$. In
other words, the confidence  is the ratio between the number of
instances when the outcome $j$ is correct and the total number of
instances when the outcome $j$ is detected.
In this paper we shall denote the maximum possible value of the confidence for the state $j$ by $C_j$, that is, 
\begin{equation}
\label{conf1}
 C_j   ={\rm max}\left\{\frac{\eta_j{\rm Tr} (\rho_j\Pi_j)}{{\rm Tr} (\rho\Pi_j)}\right\}
={\rm max}\left\{{\rm Tr}\left[\tilde{\rho}_j
   \frac{ \rho^{1/2} {\Pi}_j\, \rho^{1/2}}{{\rm Tr}
  (\rho\Pi_j)}\right]\right\},
\end{equation}
where the maximum is taken with respect to all possible measurements. 
Here we have defined the transformed density operators
\begin{equation}
\label{rho}
 \tilde{\rho}_j =\rho^{-1/2} \eta_j\rho_j\,\rho^{-1/2} \qquad {\rm
 with}\;\;  \sum_{j=1}^N
\tilde{\rho}_j = I_d.
\end{equation}
The maximum confidence  $C_j$ is equal to the largest eigenvalue of the operator $\tilde{\rho}_j$, 
and it  is obtained in a measurement where the operator  $ { \rho^{1/2} {\Pi}_j\,
\rho^{1/2}}$  has its support in the eigenspace  belonging
to the largest eigenvalue of   $\tilde{\rho}_j$ \cite{croke,herzog1}. Denoting the projector onto this eigenspace by $P_j$, 
we can write  the spectral decomposition of   $\tilde{\rho}_j$ as 
\begin{equation}
\label {rhoj}
\tilde{\rho}_j =C_jP_j +\!\! \sum_{k=m_j+1}^d
\!\!\nu^{(j)}_k|\nu_k^{(j)}\rangle\langle \nu_k^{(j)}|,
\end{equation}
 where  $m_j$ is the degree of  degeneracy of the largest eigenvalue  and  where
\begin{equation}
\label {Pj}
P_ j=\sum_{k=1}^{m_j}  |\nu_k^{(j)}\rangle\langle
\nu_k^{(j)}|, \quad \sum_{k=1}^d |\nu_k^{(j)}\rangle\langle \nu_k^{(j)}|=I_d. 
\end{equation}
The detection operators describing a maximum-confidence measurement then can be represented as \cite{herzog1} 
\begin{equation}
\label {Pij}\Pi_j = \!\!\!\sum_{k,k^{\prime}=1}^{m_j} \!\!a_{k
k^{\prime}}^{(j)} \rho^{-1/2}|\nu_k^{(j)}\rangle\langle
\nu_{k^{\prime}}^{(j)}|\rho^{-1/2}
\end{equation}
$(j=1,\ldots,N)$ with suitably chosen coefficients $a_{k
k^{\prime}}^{(j)}$.
Let us introduce the
projector $\Lambda_j$ onto the support of $\Pi_j$, that is 
\begin{equation}
\label {Lambda-j} \Lambda_j = \mbox {projector onto
span}\{\rho^{-1/2}|\nu_1^{(j)}\rangle, \ldots,
\rho^{-1/2}|\nu_{m_j}^{(j)}\rangle\}.
\end{equation}
Equation (\ref{Pij}) shows that a maximum-confidence measurement is defined  by the property that 
\begin{eqnarray}
\label {requ}    \Pi_j= \Lambda_j  \Pi_j \Lambda_j \quad (j=1,\ldots,N)
\end{eqnarray}
which restricts the supports of the detection operators $\Pi_j$ 
to the required subspaces.
We can  derive an explicit expression for the projectors $\Lambda_j$.  
From Eq. (\ref{Lambda-j}) we get $\Lambda_j \rho^{-1/2}|\nu_k^{(j)}\rangle=
\rho^{-1/2}|\nu_k^{(j)}\rangle $  for $k=1,\ldots,m_j$ 
 and thus  $\Lambda_j \rho^{-1/2}P_j= \rho^{-1/2}P_j$  which yields the relation
$ \rho^{1/2}\Lambda_j \rho^{1/2} \rho^{-1} P_j = P_j$.
The latter relation can be further modified. Taking into account  that because of  Eq. (\ref{Lambda-j})
 the projector
onto the support of  $\rho^{1/2}\Lambda_j \rho^{1/2}$ is given by $P_j$ we get
$\rho^{1/2}\Lambda_j \rho^{1/2}P_j \rho^{-1}P_j = P_j$ and finally 
\begin{equation}
\label {Lambda-j-a} \Lambda_j = \rho^{-1/2} \left(P_j
\rho^{-1}P_j\right)^{-1}\rho^{-1/2},
\end{equation}
where we used the convention that the inverse of an operator is
defined on its support, that is $\left(P_j
\rho^{-1}P_j\right)^{-1}=P_j \left(P_j \rho^{-1}P_j\right)^{-1}P_j$.
Clearly,  $\Lambda_j^2 =  \Lambda_j$, as expected for a projector.
Since the ratios in Eq. (\ref{conf1}) do not change when the
operators $\Pi_j$ $(j=1,\ldots, N)$ 
are multiplied by arbitrary constants,   it is always possible
to construct a  measurement where 
Eqs. (\ref{requ})  and  (\ref{cond1}) are fulfilled with a  suitable detection operator $\Pi_0$ .

\section{Optimized measurement}

\subsection{Necessary and sufficient optimality conditions}

In this paper we consider the measurement that
discriminates $N$ mixed quantum states with maximum confidence for
each conclusive result, and that is optimized by the additional requirement that the overall failure
probability $Q$ be as small as
possible. The latter is defined as  the overall probability of
inconclusive results, $Q= {\rm Tr}(\rho\Pi_0)$.
 It is convenient to  introduce the overall probability $R$  that the measurement
 delivers a conclusive result, no matter whether this
result is correct or wrong, given by
\begin{equation}
\label {PS1}   R= 1-Q=    \sum_{j=1}^N  {\rm
Tr}(\rho\Pi_j)=\sum_{k,j=1}^N
 \eta_k{\rm Tr}(\rho_k{\Pi}_j).
 \end{equation}
The task  is  now to determine the specific  measurement where
for each of the conclusive outcomes $j$ the confidence takes its
maximum possible value $C_j$ while  $R$ is as large as possible. For this
purpose we have to solve the
 optimization problem
\begin{equation}
\label {primal1}{\rm maximize}\,\; R  = \sum_{j=1}^N {\rm Tr}(\Lambda_j\rho
\Lambda_j\Pi_j), \;\;
 \mbox{subject to}
\;\;\Pi_0 \geq  0,
\end{equation}
 where   Eq.  (\ref{requ}) and the cyclic invariance of the trace has been used. 
 This  problem can be cast into a different form by
observing that the relation 
\begin{equation}
\label {dual2a} {\rm Tr} Z - R= {\rm Tr }(Z\Pi_0) + \sum_{j=1}^N
{\rm Tr }[\Lambda_j(Z-\rho)\Lambda_j\Pi_j]
\end{equation}
is identically fulfilled for any operator $Z$, as can be seen 
   after  replacing $Z$ on the left-hand side of  Eq. (\ref{dual2a}) by $ZI_d$ with  $I_d=\Pi_0+ \sum_j\Lambda_j\Pi_j\Lambda_j$.
Since the detection operators are positive,  we conclude from Eq. (\ref{dual2a}) that the inequality 
\begin{equation}
\label {dual2} 
{\rm Tr} Z - R \geq 0
\end{equation}
holds true provided that  the  sufficient positivity conditions
\begin{eqnarray}
\label {Z1}
Z\geq 0,\quad  \Lambda_j(Z-\rho)\Lambda_j \geq 0\;\;\;(j=1,\ldots,N) 
\end{eqnarray}
are satisfied.  Equation (\ref{dual2}) implies that the
discrimination probability $R$ cannot be larger than the smallest
possible value of ${\rm Tr} Z$. In other words,
 the
minimum of ${\rm Tr} Z$ under the constraints given by Eq. (\ref{Z1})
determines an upper bound for $R$. In order to determine this bound,
we thus arrive at the alternative optimization problem
\begin{eqnarray}
\label {dual3} && {\rm minimize} \;\, {\rm Tr} Z, \;\;
\mbox{subject to Eq. (\ref{Z1})}.\qquad
\end{eqnarray}
When  the bound is reached, that is when $R= {\rm min}({\rm Tr}\, Z)$,
 the right-hand side of Eq. (\ref{dual2a}) vanishes
 for the optimum  operator  $Z$  solving the minimization problem
given in  Eq.  (\ref{dual3}).  Due to the
positivity conditions in  Eq. (\ref {Z1}) and the positivity of the detection
operators, all traces in Eq.
(\ref{dual2a}) are taken over positive operators. Hence the
expression on the right-hand side of  Eq. (\ref{dual2a}) can only vanish when the conditions 
\begin{eqnarray}
\label {Z3}
 Z \Pi_0 = 0,\quad \Lambda_j(Z-\rho)\Pi_j =0\;\;\;(j=1,\ldots,N)  
\end{eqnarray}
are fulfilled, where in the second equation again Eq. (\ref{requ}) has been used.  Since the converse is obvious,
Eq. (\ref {Z3}) is necessary and sufficient for optimality, 
provided that
 the  positivity conditions in Eq. (\ref{Z1}) are fulfilled.
 As will be shown in the Appendix by a proof which is analogous to a recent proof concerning  minimum-error discrimination  \cite{barnett-croke}, the conditions in Eq. (\ref{Z1}) are not only
 sufficient, but also necessary for the validity of Eq. (\ref{dual2}). Together  
 Eqs.  (\ref{Z1}) and  (\ref{Z3}) then represent necessary and sufficient optimality conditions. When we can find an operator $Z$
and positive detection operators  $\Pi_0$ and $\Pi_j= \Lambda_j  \Pi_j \Lambda_j$ $(j=1,\ldots,N)$ which satisfy these conditions, then the 
detection operators determine the optimum measurement, that is the maximum-confidence measurement which maximizes $R$, or minimizes the failure probability $Q$, respectively. 

The results  can be written in an alternative way by taking into account that  
\begin{equation}
\label {sigma2}
\Lambda_j \eta_j\rho_j \Lambda_j= C_j
\Lambda_j\,\rho\,\Lambda_j
\end{equation}
which follows from  Eqs. (\ref {rho}), (\ref {rhoj}) and
(\ref{Lambda-j-a})  using $P_j  \tilde{\rho}_j P_j=C_jP_j  \rho^{-1/2}\rho  \rho^{-1/2}P_j$.
We thus arrive at the necessary and sufficient optimality conditions  
\begin{eqnarray}
\label {sigma1a}
\! \Lambda_j(Z-\rho)\Lambda_j\!\! &=&\!\! \Lambda_j  \!\left( \!Z- \frac{\eta_j\rho_j}{C_j}\!\right) \!\Lambda_j \!\geq\! 0,\;\;\,  Z\geq 0,\\\label {sigma1b}
\! \Lambda_j(Z-\rho)\Pi_j \!\! &=& \!\! \Lambda_j  \!\left( \!Z -\frac{\eta_j\rho_j}{C_j} \!\right)\!\Pi_j\! =\!0, \;\; Z \Pi_0\! =\!  0, \quad \quad
\end{eqnarray}
with $\Pi_j=\Lambda_j\Pi_j\Lambda_j$ and  $j=1,\ldots,N$. The rate of conclusive results reads   
\begin{equation}
\label {sigma4} 
R=1-Q= \sum_{j=1}^N {\rm Tr}(\rho\Pi_j)=
 \sum_{j=1}^N \frac{\eta_j}{C_j}\;{\rm  Tr}(\rho_j\Pi_j).
\end{equation}
When $C_j=1$ for $j=1,\ldots,N$ Eqs. (\ref{sigma1a}) and (\ref{sigma1b}) coincide with the conditions 
for optimum unambiguous discrimination that have been derived 
by Eldar {\it et. al}.  \cite{eldar} from a semidefinite programming problem using duality theory in linear optimization.
In this context, the optimization
problem posed in Eq. (\ref{primal1}) is called the primal problem,
while the problem posed in Eq. (\ref{dual3}) is denoted as the dual
problem. The specialization to unambiguous discrimination will be discussed  in more detail in Sec. III.D.

\subsection  {Rank of the optimum failure operator}

From the optimality conditions we can draw a conclusion about the
rank of the failure operator in the optimum measurement for
maximum-confidence discrimination.  Equation (\ref{Z3})  implies that for
$Z\neq 0$ the operator $\Pi_0$ is orthogonal to $Z$ which means that
the eigenstates
 of $\Pi_0$ cannot span the full $d$-dimensional Hilbert space ${\cal H}_d$ introduced
 by means of Eq. (\ref{rho-x}). More precisely, the first equality in Eq. (\ref{Z3}) requires that 
\begin{eqnarray}
\label{rank}
 {\rm rank}\;Z + {\rm rank}\;\Pi_0 \leq d
\end{eqnarray}
since otherwise the joint eigensystems of $Z$ and $\Pi_0$ would
consist of more than $d$ states in ${\cal{H}}_d$ and
due to the linear dependence of these states the operators would not
be orthogonal.
 On the other hand,
the second positivity constraint in Eq. (\ref{Z1}) implies 
that the relation $  {\rm
rank}(\Lambda_j Z \Lambda_j) \geq   {\rm rank}(\Lambda_j\rho\Lambda_j )$
is fulfilled for
each value of $j$.  Since  obviously ${\rm rank}\;Z \geq  {\rm
rank}(\Lambda_j Z \Lambda_j)$ we arrive at
\begin{equation}
\label{rank1} {\rm rank}\;Z \geq {\rm max}_j \{{\rm
rank}(\Lambda_j\rho\Lambda_j )\}= {\rm max}_j \{{\rm
rank}(\Lambda_j\rho_j\Lambda_j )\}
\end{equation}
where for the last equality sign Eq. (\ref{sigma2}) has been taken into account. Combining Eqs.
(\ref{rank}) and (\ref{rank1}) we obtain the condition
\begin{eqnarray}
\label {rank3}  {\rm rank}\;\Pi_0 \leq d-{\rm max}_j \{{\rm
rank}(\Lambda_j\rho_j\Lambda_j )\}
\end{eqnarray}
which restricts the rank of the failure operator in a measurement
performing maximum-confidence discrimination with minimum failure
probability.

\subsection{Dimensionality of the optimization problem}

We introduce  the $d^{\prime}$-dimensional subspace ${\cal
H}_{d^{\prime}}$  jointly spanned by
 the projectors $\Lambda_1,\ldots, \Lambda_N$. When we define
\begin{equation}
\label {Lambda-prime}
  \rho^{\prime}=\frac{\Lambda\rho\Lambda}{{\rm Tr}(\rho\Lambda)},\quad \Lambda = \mbox{projector onto span}
(\Lambda_1,  \ldots \Lambda_N), 
\end{equation}
the optimization problem posed  in  Eq. (\ref{primal1}) corresponds to the maximization of
\begin{equation}
\label{PS-d}    R={\rm Tr}(\rho\Lambda)  R^{\prime} \quad {\rm
with}\;\; R^{\prime} =\sum_{j=1}^N {\rm
Tr}(\rho^{\prime}\Lambda_j\Pi_j\Lambda_j),
\end{equation}
under the constraint that  $\Pi_0= I_d- \Lambda +\Pi_0^{\prime} \geq
0$. Here we introduced  the operator $ \Pi_0^{\prime}=
\Lambda- \sum_{j=1}^N\Lambda_j\Pi_j\Lambda_j$ which has its support
in ${\cal H}_{d^{\prime}}$. Since  $I_d-\Lambda$ is the projector onto the subspace which is 
orthogonal to ${\cal H}_{d^{\prime}}$,  the positivity constraint on $\Pi_0$  
is satisfied provided that $\Pi_0^{\prime} \geq 0$. 
The  optimization is  thus reduced  to
the maximization of  $R^{\prime}$ subject to 
$\Pi_0^{\prime} \geq 0$, which is a problem in a Hilbert space of 
 dimension  $d^{\prime} \leq d$. The latter problem is 
formally equivalent to the original optimization problem when we identify
$\Lambda$ with the identity operator in ${\cal H}_{d^{\prime}}$. Hence 
 the condition
\begin{equation}
\label {standard}   \mbox{span} (\Lambda_1,  \ldots \Lambda_N) =
{\cal H}_d, \quad \Lambda=I_d
\end{equation}
 defines a standard form of the optimization problem to which the general problem can be
reduced with the help of Eqs. (\ref{Lambda-prime}) and  (\ref{PS-d}).  It is therefore sufficient to restrict the investigations  to 
 the  case $d^{\prime}= d$.
Eq. (\ref{PS-d}) shows that  for  $d^{\prime} < d$ maximum-confidence discrimination without inconclusive results  is never possible since in this case ${\rm Tr}(\rho\Lambda) < 1$, which yields a rate of conclusive results $R< 1$ even when $R^{\prime}=1$, that is, even when $ \Pi_0^{\prime}=0$.

\subsection{Specialization to unambiguous discrimination}

When each of the given states can be discriminated with perfect
confidence, that is for 
\begin{eqnarray}
\label {unamb0} 
C_1=\dots =C_N=1,
\end{eqnarray}
 the maximum-confidence measurement is equivalent to a measurement
which unambiguously discriminates between the $N$  states. Indeed,
from Eqs. (\ref{rho}) and (\ref{rhoj}) we obtain the relation  $P_j  \tilde{\rho}_j=C_jP_j \sum_{k=1}^N \tilde{\rho}_k$. If $C_j=1$ this relation can be only fulfilled when $P_j\tilde{\rho}_k = 0$ for any
$k$ with $k\neq j$ and when  therefore also  $P_jP_k = 0$
which follows from representing the positive operator $\tilde{\rho}_k$ by Eq. (\ref{rhoj}). 
This means that the projectors onto the eigenspaces belonging to the largest eigenvalues of  $\tilde{\rho}_j$ and
$\tilde{\rho}_k$, respectively,  have to be  mutually orthogonal.   
  Equation (\ref{unamb0}) therefore  necessarily requires that for $k,j=1,\ldots ,N$
\begin{eqnarray}
\label {unamb1}  P_kP_j=\delta_{kj},\quad  \Lambda_k\rho_j =  \Pi_k\rho_j=0  \quad {\rm for}\;\; k\neq j,
\end{eqnarray}
 where the second relation follows from the first with the help of Eqs. (\ref{requ}) and (\ref{Lambda-j-a}). The last  equality represents the condition for
unambiguous discrimination.
 Conversely, Eq. (\ref{conf1}) shows immediately that Eq. (\ref{unamb0}) follows from Eq. (\ref{unamb1}).

We emphasize that when the maximum confidence is equal to unity for some of
the states and smaller for the rest of them, the strategies of maximum-confidence
discrimination and of unambiguous discrimination are different.
In unambiguos discrimination  the detection
operator $\Pi_j$ is zero for a state where $C_j<1$  since this state cannot be unambiguously discriminated and therefore always yields an inconclusive result.  Examples to
elucidate this difference for $N=2$ are presented in \cite{herzog1}.

Let us specialize the previous  considerations  about a standard form of the optimization problem to the case where Eq. (\ref{unamb0}) holds.    
From Eqs. (\ref{unamb1}) and  (\ref{rhoj})  it follows that $P= \sum_{j=1}^N P_j$
 projects onto a  subspace of dimension  $d^{\prime}=\sum_{j=1}^N m_j$.  If $d^{\prime}=d$,  $P$ is equal to the identity operator $I_d$ which means that  also $\Lambda=I_d$ since $P_j$ is the support of $\Lambda_j$. Using  Eq. (\ref{rhoj}) with $C_j=1$ in order to calculate  $\sum_{j=1}^N \tilde{\rho}_j$ and comparing the result with the resolution of the identity in Eq.  (\ref{rho}) it  follows that $ \tilde{\rho}_j= P_j$ ($j=1,\ldots,N$). This
implies that for $\Lambda=I_d$ there does not exist a common subspace for the supports
of any two of the operators $\tilde{\rho}_j$ and hence also no
common subspace for any two of the given density operators
${\rho}_j$. 
Hence the reduction of the optimization problem to the
standard form, characterized by Eq. (\ref{standard}), is equivalent to the
elimination of common subspaces between the given density operators.
Our treatment provides a recipe how to perform this elimination, thus
extending the corresponding reduction theorem established for the
optimum unambiguous discrimination of two mixed states  \cite{raynal} to an
arbitrary number $N$ of mixed states.

\subsection{Maximum-confidence discrimination for $N=2$}

The case  $N=2$ is exceptional since in this case 
the relation $P_1P_2=0$,
which according to Eq. (\ref{unamb1})  holds  when $C_1=C_2=1$,  that is for unambiguous discrimination, remains 
valid for any maximum-confidence measurement,  with arbitrary values $C_1$ and $C_2$.
This is due to the fact that the transformed
density operators $\tilde{\rho}_1$ and $\tilde{\rho}_2 = I_d- \tilde{\rho}_1$
 in Eq. (\ref{rhoj}) have the same system of eigenstates. 
 The eigenstates of
$\tilde{\rho}_1$ belonging to its largest eigenvalue
are  associated with the
smallest eigenvalue of $\tilde{\rho}_2$,  and vice versa \cite{herzog1},  and the projectors $P_1$ and $P_2$ are therefore orthogonal.  
 
When $P_1+P_2 = I_d$, that is when the optimization problem has been reduced to the standard form characterized by  Eq. (\ref{standard}),  we obtain from Eq. (\ref{rhoj}) 
\begin{equation}
\label {N2a} \tilde{\rho}_1  =  C_1P_1 + (1-C_2) P_2,\quad
\tilde{\rho}_2  =  C_2P_2 + (1-C_1) P_1.
 \end{equation}
Let us  introduce the operators   $\sigma_j=\rho^{1/2}P_j \rho^{1/2}$ $(j=1,2)$ with $\sigma_1+\sigma_2=\rho$. Making use of Eq. (\ref {Lambda-j-a}), we find that   
\begin{equation}
\label {N2b} \sigma_1\Lambda_2\! =\! \sigma_2\Lambda_1 \!= \!0, \quad\Lambda_j\rho\Lambda_j= \Lambda_j\sigma_j\Lambda_j=
  \Lambda_j\sigma_j^{\prime}\Lambda_j{\rm Tr}\sigma_j,
 \end{equation}
 where the operators $\sigma_j^{\prime}=({\rm Tr} \sigma_j )^{-1} \sigma_j$ can be interpreted as density operators.  
The first equation implies that $\sigma_1^{\prime}\Pi_2 = \sigma_2^{\prime}\Pi_1=0$,  due to Eq. (\ref{requ}). 
Hence after substituting the second equation into the optimality conditions, Eqs. (\ref{sigma1a}) and  (\ref{sigma1b}), we find  that the conditions for the optimized maximum-confidence discrimination between $\rho_1$ and $\rho_2$  are equivalent to the  conditions for the optimum unambiguous discrimination between   $\sigma_1^{\prime}$ and $\sigma_2^{\prime}$, occurring with the prior probabilities ${\rm Tr}\sigma_1$ and  ${\rm Tr}\sigma_2$, respectively. The results  obtained from studying the latter problem \cite{rudolph,raynal,eldar,HB,BFH,herzog,raynal2,kleinmann} are therefore directly applicable to the former.  
In explicit terms, the operators $\sigma_j=\rho^{1/2}P_j \rho^{1/2}$
read     
\begin{equation}
\label {N2c} \sigma_1\! =\! \frac{\eta_1\rho_1 C_2 \!- \!
\eta_2\rho_2  (1 \!- \!C_1)}{C_1\!+\! C_2-1},\;\;  \sigma_2\! =\!
\frac{\eta_2\rho_2 C_1\!-\! \eta_1\rho_1 (1\!-\!C_2)}{C_1\!+\!
C_2-1},
 \end{equation}
as follows from  Eqs. (\ref{N2a}) and (\ref{rho}).

 \section{Application to symmetric states}

\subsection{Properties of the optimum measurement}

In the following we assume that 
 the $N$ states occur with equal prior probability $1/N$,
and that  they are cyclically symmetric in the sense that
neighboring states arise from each other by the same unitary
transformation. The density operators are given as  
\begin{equation}
\label {s1} \rho_{j+1}= V^{j}\rho_1 V^{\dag j } \qquad {\rm with}
\quad V^{\dag}V= V^{N}=I_d\end{equation}
for $j=0,\ldots,N-1$, where without lack of generality  $\rho_1$ has been taken as the reference operator.
By re-numbering the states it can be easily seen that \cite{ban}
\begin{eqnarray}
\label {s3} \rho=\frac{1}{N}\sum_{j=1}^N \rho_j =V \rho V^{\dag},
\qquad [V,\rho]=0,
\end{eqnarray}
where the second equation follows from the first. Since the Hermitian operator $\rho$ and the unitary operator $V$ commute  the two operators  have the same eigenbasis in ${\cal H}_d$. 
Upon denoting their eigenstates by $\{|r_l\rangle\}$ with
$l=1,\ldots,d$, in correspondence with Eq. (\ref{rho-x}), we get from
Eq. (\ref{s1}) the spectral representation
\begin{equation}
\label{s3a} V\!= \! \sum_{l=1}^d v_l|r_l\rangle\langle r_l| \qquad
{\rm with}\quad \sum_{j=1}^{N} (v_lv_{l^{\prime}}^{\ast})^{j} = N
\delta_{ll^{\prime}}
\end{equation}
and with $|v_l|^2=v_l^N=1$,  yielding 
\begin{equation}
\label{s3b} \langle r_l | \rho_{j}|r_l\rangle =\langle
 r_l|\rho|r_l\rangle \equiv r_l \quad (j=1,\ldots, N).
\end{equation}
This shows that in the eigenbasis of the symmetry operator the
 symmetric  density operators have the same diagonal elements.
 The second equality in Eq. (\ref{s3a}) follows from expanding 
 the density operator $\rho_1$
in terms of the eigenbasis of
 $\rho$  or $V$, respectively, arriving at
\begin{equation}
\label {s3c} \rho\!  = \! \frac{1}{N}\sum_{j=1}^{N} V^{j}\rho_1
V^{\dag j} = \!\sum_{l,l^{\prime}=1}^d \!\langle
r_l|\rho_1|r_{l^{\prime}}\rangle \! \sum_{j=1}^{N}\frac{
(v_lv_{l^{\prime}}^{\ast})^{j}}{N} |r_l\rangle\langle
r_{l^{\prime}}|.
\end{equation}
Taking Eq. (\ref{s3b}) into account, the desired equality is immediately obvious by
comparison with Eq. (\ref{rho-x}).

After these general considerations about symmetric states we now
focus on their maximum-confidence discrimination. Using  $[V,\rho]=0$ we find from Eqs. (\ref{rho}) and (\ref{s1})  that for symmetric states the transformed density
operators obey the equation
\begin{equation}
\label {s4}  \tilde{\rho}_{j+1}= \frac{1}{N}\rho^{-1/2}\rho_{j+1}
\rho^{-1/2}=  V^{j}\tilde{\rho}_1 V^{\dag j}
\end{equation}
with $j=0,\ldots, N-1$. Clearly, the eigenvalue spectrum is the same
for each of the states $\tilde{\rho}_j$.   As a consequence, also
the maximum confidence for each of the outcomes, being equal
to the largest eigenvalues of the transformed operators
$\tilde{\rho}_j$, is  the same,
\begin{eqnarray}
\label {s5} C_1= \dots = C_N\equiv C.
\end{eqnarray}

We mention at this point that whenever in a discrimination measurement the confidence has the same value $C$ for each of the outcomes,  then in  this measurement  the relation 
$\eta_j{\rm Tr} (\rho_j\Pi_j)= C {\rm Tr} (\rho\Pi_j)$  is fulfilled for $j=1,\ldots,N$, as becomes obvious from the definition of the confidence in the text  before Eq. (\ref{conf1}).   Summation over all states on both sides of the latter equation yields   
\begin{equation}
\label {s5a} P_{corr}=\sum_{j=1}^N \eta_j{\rm Tr} (\rho_j\Pi_j)=C \sum_{j=1}^N{\rm Tr} (\rho\Pi_j)= C(1-Q)
\end{equation}
where $P_{corr}$ is the overall probability of getting a correct result. 

Due to Eq. (\ref{s4}) the projectors $P_j$ onto the
eigenspaces belonging to the largest eigenvalue of the operators
$\tilde{\rho}_j$, see Eq. (\ref{rhoj}),  obey the same symmetry as the given density
operators. 
Since $\rho$ and $V$ commute, this symmetry is conveyed
to the projectors $\Lambda_j$ defined in Eq. (\ref{Lambda-j}) which
specify the supports of the detection operators $\Pi_j$. We thus get
\begin{equation}
\label {s7} 
P_{j+1}= V^{j} P_1 V^{\dag j},
\quad \Lambda_{j+1}= V^{j} \Lambda_1 V^{\dag j}.
\end{equation}
The optimized
maximum-confidence  measurement, minimizing the failure probability,
is determined by the optimal  detection operators satisfying Eqs. (\ref{Z1}) and  (\ref{Z3}). 
 Let
us assume that $\Pi_1$ is an element of the set of optimal detection
operators and consider the operators
\begin{equation}
\label {s7b} \Pi_{j+1}= V^{j} \Pi_1 V^{\dag j}
\end{equation}
with  $j=0,\ldots, N-1$. By the same arguments that led to the derivation of the
commutation relation in   Eq. (\ref{s3}) we find that the operator
$\sum_{j=1}^N\Pi_j= I-\Pi_0$ commutes with $V$. This implies
\begin{equation}
\label {s7c}  [\Pi_0,V]=  [\Pi_0,\rho]=  0, \quad [Z,V]=  [Z,\rho]=0,
\end{equation}
where the second relation follows from the first when we  take into account that  $[Z,\Pi_0]=0$,
due to Eq. (\ref{Z3}) and the hermiticity of the operators $Z$
and $\Pi_0$. Using Eqs. (\ref{s7}) --  (\ref{s7c}) it follows that  for $j=0,\ldots, N-1$   
\begin{equation}
\label {s7c1} \Lambda_{j+1}(Z-\rho)\Lambda_{j+1}= V^{j} \Lambda_1 (
Z-\rho)\Lambda_1 V^{\dag j}.
\end{equation}
Because of Eq. (\ref{requ}) this means that if  $\Pi_1$ fulfills the optimality conditions,
then these conditions are fulfilled by any of the
operators $\Pi_j$ defined in Eq. (\ref{s7b}).  Hence 
the detection operators for the optimum measurement
can always be chosen in the form of Eq. (\ref{s7b}), that is in a form
where they have the same symmetry as the density operators.
The optimality conditions therefore reduce to Eq. (\ref{s7b}) together with  
\begin{eqnarray}
\label {sigma1as}
\! \Lambda_1(Z-\rho)\Lambda_1\!\! &=&\!\! \Lambda_1 
 \!\left( \!Z- \frac{\rho_1}{N\;C }\!\right) \!\Lambda_1 \!\geq\! 0,\;\;\,  Z\geq 0,\\
\label {sigma1bs}
\! \Lambda_1(Z-\rho)\Pi_1 \!\! &=& \!\! \Lambda_1
  \!\left( \!Z -\frac{\rho_1}{N\;C} \!\right)\!\Pi_1\! =\!0, \;\; Z \Pi_0\! =\!  0, \quad \quad
\end{eqnarray}
where we used   Eqs. (\ref{sigma1a}) and  (\ref{sigma1b}). We mention that for the case of optimum  unambiguous
discrimination  of symmetric states the corresponding  symmetry property of the
optimum detection operators was derived by Chefles and Barnett \cite{chefles-symm} for
pure states and by Eldar {\it et al.} \cite{eldar} for mixed states. From Eqs.
(\ref{s7b}) and  (\ref{s7c}) it follows that in the optimum  measurement  
the probability $R$ of conclusive results,  defined in  Eq. (\ref{PS1}), can be written as 
\begin{equation}
\label {s7d} R = 1-Q=N {\rm Tr}(\rho\Pi_1 )= \sum_{l=1}^{d} r_l N\langle
r_l|\Pi_1|r_l\rangle,
\end{equation}
where the spectral representation of $\rho$ has been used.  Applying
the properties of the symmetry operator $V$, given in  Eq.
(\ref{s3a}),  we arrive at the failure operator
\begin{equation}
\label {s7e} \Pi_0= I- \sum_{j=1}^{N} V^{j}\Pi_1 V^{\dag j}=
\sum_{l=1}^{d}\left(1-N\langle
r_l|\Pi_1|r_l\rangle\right)|r_l\rangle\langle r_l|.
\end{equation}
The optimum detection operator $\Pi_1$ maximizes $R$ under the constraint that $\Pi_0 \geq 0$ and obeys the condition $\Pi_1 = \Lambda_1\Pi_1 \Lambda_1$, which guarantees
maximum-confidence discrimination.

According to Eq. (\ref{Lambda-j-a}) the projector $\Lambda_1$ is
determined by the projector $P_1$ which in turn, together with the
maximum confidence $C,$  follows from the
spectral decomposition of $\tilde{\rho}_1= N^{-1}\rho^{-1/2}\rho_{1}
\rho^{-1/2}$.   After expanding $\tilde{\rho}_1$ in terms of the
eigenstates of $\rho$ or $V$, respectively,  Eq. (\ref{rhoj}) takes
the form
\begin{equation}
\label {s6} \tilde{\rho}_1 = \!\!\sum_{l,l^{\prime}=1}^d
\frac{\langle r_l|\rho_1|r_{l^{\prime}}\rangle
}{N\sqrt{r_lr_{l^{\prime}}}} \,|r_l\rangle\langle r_{l^{\prime}}| =
CP_1+ \!\!\sum_{k=m+1}^d \!\!\nu_k^{(1)}|\nu_k^{(1)}\rangle\langle \nu_k^{(1)}|
\end{equation}
with $P_1 = \sum_{k=1}^{m} |\nu_k^{(1)}\rangle\langle \nu_k^{(1)}| $. Here
$m={\rm rank}\,P_1$ is the degeneracy of the largest eigenvalue,
$C$.
Clearly, the rank of $P_1$ depends on the
 matrix elements $ \langle
r_l|\rho_1|r_{l^{\prime}}\rangle$, that is on the representation of $\rho_1$ in the eigenbasis of the symmetry
operator.

\subsection{General solution for one-dimensional detection operators}

In the rest of the paper we suppose that  the largest eigenvalue of
the transformed density operator $\tilde{\rho}_1$ is nondegenerate,
that is $m=1$ in Eq. (\ref{s6}). Provided that the spectral representations of $\rho$ and $\tilde{\rho}_1$ are known,  in this simple case  the optimization problem posed in Eq. (\ref{primal1})  can be readily solved in a direct calculation, 
 without resorting to the general optimality conditions.  For convenience we drop the superscript which indicates 
the number of the state, that is we use the notation $|\nu_1^{(1)}\rangle=
|\nu_1\rangle$. With the help of  Eq. (\ref{Lambda-j-a}) we
then get
\begin{eqnarray}
\label {r3} P_{1}= |\nu_1\rangle\langle \nu_1|, \quad
\Lambda_1=\rho^{-1/2}\frac{|\nu_1\rangle\langle \nu_1|}{\langle
\nu_1|\rho^{-1}|\nu_1\rangle}\rho^{-1/2},
\end{eqnarray}
 where $|\nu_1\rangle$ is the eigenstate of $\tilde{\rho}_1$ belonging to its largest
 eigenvalue.
The requirement $\Pi_1= \Lambda_1\Pi_1\Lambda_1$ guaranteeing
maximum-confidence discrimination leads to the  Ansatz
\begin{equation}
\label {r4} 
\Pi_1=\alpha\,\rho^{-1/2}|\nu_1\rangle\langle \nu_1|\rho^{-1/2}
\end{equation}
which because of  Eqs.  (\ref{s7d})  and  (\ref {s7e}) yields
\begin{equation}
\label {r5} R= N\alpha,\quad  \Pi_0= \sum_{l=1}^{d}\left(1-N\alpha \frac{
|\langle r_l|\nu_1\rangle|^2} {r_l} \right)|r_l\rangle\langle r_l|.
\end{equation}
In order to maximize $R$ we have  to find
the largest admissible  value of $\alpha$ which is consistent with the
constraint $\Pi_0\geq 0$. Clearly,  this constraint requires that
$\alpha N |\langle r_l|\nu_1\rangle| ^2\leq r_l $ for each value of $l$.
The largest possible value of $\alpha$  and the resulting minimum failure
probability $Q=1-R$  are therefore
\begin{equation}
\label {r5a}  \alpha_{opt}= \frac{1}{N}{\rm
Min}_l \left \{\frac{r_l}{ |\langle
r_l|\nu_1\rangle|^2}\right \},\quad Q_{min} = 1- N\alpha_{opt},
\end{equation}
where the minimum is taken with respect to the different values of $l$.

In general, there can exist $l_0$ different values of $l$   for
which the fraction $r_l/ |\langle r_l|\nu_1\rangle|^2  $ takes the
same minimal value, where $1\leq l_0 \leq d$. Let us number the
eigenstates $|r_l\rangle$ in such a way that the condition
$r_l=\alpha_{opt}N|\langle r_l|\nu_1\rangle|^2$ is fulfilled for
$l=1,\ldots, l_0$, yielding the failure operator
\begin{equation}
\label {r6} \Pi_0 = \!\sum_{l=l_0+1}^d\!\left(1- N \alpha_{opt}\frac{
|\langle r_l|\nu_1\rangle|^2}{r_l}\right)|r_l\rangle\langle r_{l}|\;
\end{equation}
with ${\rm rank}\,\Pi_0 = d- l_0$. Clearly, ${\rm rank}\, \Pi_0 \leq
d-1,$ in accordance with Eq. (\ref{rank3}). For $l_0=d$ we get
$\Pi_0=0$ which means that  in this special case inconclusive results do not occur in the
optimized maximum-confidence measurement. This case arises when
$r_l=|\langle r_l|\nu_1\rangle|^2 $ for each value of $l$, that is
when $\rho=|\nu_1\rangle \langle \nu_1|$. An example will be treated in Sec. IV.C.

It is easy to verify that the solution
fulfills the
necessary and sufficient optimality conditions. 
 For this purpose we introduce the operator 
\begin{equation}
\label {r7} Z =  \frac{N\alpha_{opt}}{l_0}\sum_{l=1}^{l_0}
|r_l\rangle\langle r_l|= \frac{1}{l_0}\sum_{l=1}^{l_0}
\frac{r_l}{|\langle r_l|\nu_1\rangle|^2}|r_l\rangle\langle r_l|
\end{equation}
which is obviously  positive and orthogonal to $
\Pi_0$, that is  $Z\geq 0$ and $Z\Pi_0 = 0$. Moreover, we obtain
\begin{equation}
\label {r8} \langle \nu_1 |\rho^{-1/2} (Z-\rho)
\rho^{-1/2}|\nu_1\rangle =0
\end{equation}
which because of Eq. (\ref{r3})  means that
$\Lambda_1(Z-\rho)\Lambda_1=0$ and therefore also
$\Lambda_1(Z-\rho)\Pi_1=0$. Hence  Eqs. (\ref{sigma1as}) and  (\ref{sigma1bs}) are satisfied.

\subsection{Examples}

In our examples we consider  the discrimination of $N$
equiprobable symmetric mixed states of the form
\begin{equation}
\label {r10} \rho_{j}= p\,|\psi_j\rangle\langle \psi_j|+
\frac{1-p}{d}\,I_d \quad(j=1,\ldots,N),
\end{equation}
where $|\psi_j\rangle$ is normalized to unity and where the parameter $p$ with $0\leq p \leq 1$ is related to the purity of
the given states. 
The symmetry of the set of mixed states $\rho_j$ requires that 
\begin{equation}
\label {r10a} |\psi_{j+1}\rangle = V^{j}|\psi_1\rangle \quad {\rm
with} \quad V^{\dag}V= V^{N}=I_d.
\end{equation}
  If  $N < d$ the $N$ pure
states $|\psi_j\rangle$ span a Hilbert space of dimension
$d^{\prime}$ with  $d^{\prime} < d$ and the optimization problem can be
reduced to state discrimination within the $d^{\prime}$-dimensional
Hilbert space.
Therefore without lack of generality we assume that
 the $N$  states $|\psi_j\rangle$ span the full $d$-dimensional Hilbert space.
This means that $N \geq d$ and that the expansion of
$|\psi_1\rangle$ with respect to the eigenbasis of $V$ reads
\begin{equation}
\label {r11a} |\psi_1 \rangle = \sum_{l=1}^d c_l |r_l\rangle\quad
\mbox{with}\;\;
  c_l\neq 0 \;\;\mbox{for}\;\; l=1,\dots d.
\end{equation}
 Making use of  Eqs.( \ref{s1}) and  (\ref{s3a})  we arrive at
\begin{equation}
\label {r11} \rho=\sum_{j=1}^N \frac{\rho_j}{N}=\sum_{l=1}^d
r_l|r_l\rangle\langle r_l|
 \quad
\mbox {with} \;\; r_l = p\,|c_l|^2+\frac{1-p}{d}.
\end{equation}
The general expression in Eq. (\ref{s6}) then takes the form
\begin{equation}
\label {r12} \tilde{\rho}_1=\rho^{-1/2}\frac{\rho_1}{N} \rho^{-1/2}=
\frac{p}{N}\!\sum_{l, l^{\prime}\atop (l\neq
l^{\prime})}^d \!\frac{c_l c_{l^{\prime}}^{\ast}}{\sqrt{r_l
r_{l^{\prime}} }} |r_l\rangle\langle r_{l^{\prime}}|+  \frac{I_d}{N} .
\end{equation}
In order to determine  the optimum measurement, we need to find the
spectral decomposition of $\tilde{\rho}_1$. In the following we
restrict ourselves to simple cases where this task can be solved analytically.

\subsubsection{$N$ symmetric pure states in a $d$-dimensional joint Hilbert space ($N\geq d$)}

First we treat the case that in Eq. (\ref{r10})  $p=1$  which means that  the states
to be discriminated are the $N$  equiprobable symmetric pure qudit states   $|\psi_1\rangle \dots
|\psi_N\rangle$. Clearly, when  $N >d$ the states  are linearly
dependent.
After substituting $p=1$ and $\sqrt{r_l}=
|c_l|$,  the operator
$\tilde{\rho}_1$
 in Eq. (\ref{r12}) takes the form 
\begin{equation}
\label {r13} \tilde{\rho}_1 \! =\! \frac{d}{N}|\nu_1\rangle\langle
\nu_1| \quad{\rm with}\;\;  |\nu_1\rangle\! =\!
\frac{1}{\sqrt{d}}\sum_{l=1}^d \frac{c_l}{|c_l|} |r_l\rangle \!=\! \frac{\rho^{-1/2}}{\sqrt{d}}|\psi_1\rangle.
\end{equation}
From  Eq. (\ref{rhoj}) we  obtain 
the maximum confidence $C$, and  Eq. (\ref{r5a}) with $r_l=|c_l|^2$ yields $\alpha_{opt}$ and the minimum
 failure probability $Q_{min}$. We thus arrive at  
\begin{eqnarray}
\label {r14} C = \frac{d}{N}, \qquad Q_{min} = 1-d \,{\rm
min}_l \{|c_l|^2\}.
\end{eqnarray}
Taking   Eqs. (\ref{r4}), (\ref{r13}) and  (\ref{s7b}) into account, the optimum detection operators can be represented as 
\begin{equation}
 \label {r14a}
\Pi_j= \frac{{\rm min}_l \{|c_l|^2\} }{N}
\rho^{-1}|\psi_j\rangle\langle\psi_j|\rho^{-1}\quad(j=1,\ldots,N),
\end{equation}
where $\rho=\frac{1}{N}\sum_{j=1}^N |\psi_j\rangle\langle\psi_j|$.
The
maximum confidence is solely determined by the dimension $d$ of the
Hilbert space spanned by the states and does not depend on their special form.   
For $N=3$ and $d=2$ the expression for the maximum confidence is in accordance
with the result obtained in \cite{croke}.

When  $N=d$, 
 that is when the pure states are linearly independent and  
 $C=1$, the expression for   $Q_{min}$ in Eq. (\ref{r14}) coincides with the minimum failure probability necessary for the unambiguous
discrimination of $N$ linearly independent symmetric pure states, derived by Chefles and Barnett
\cite{chefles-symm}. 
Indeed, since for these states the eigenvalues of the symmetry operator  can be
represented as $v_l={\rm exp}\left(2\pi i\frac{l}{N}\right)$
\cite{chefles-symm} we find from Eqs. (\ref{r10a}) and  (\ref{r11})
with $r_l=|c_l|^2$ that 
\begin{equation}
\label {r17} \langle \psi_j|\rho^{-1}|\psi_k\rangle = \sum_{l=1}^d
v_l^{\ast j}\frac{|c_l|^2}{r_l}  v_l^k
= \sum_{l=1}^N \left(e^{2\pi i\frac{j-k}{N}}\right)^l=N \delta_{jk},
\end{equation}
yielding due to  Eq. (\ref{r14a})  $\Pi_j|\psi_k\rangle =0$ for
$j\neq k$, which is the condition for unambiguous discrimination.
When
 $|c_l|=1/\sqrt{d}$ for $d=N$, the $N$ given states are mutually
orthogonal and $\rho=I_d/d$. The optimum  measurement is then projective with $\sum_{j=1}^N \Pi_j= I_d$ following from Eq.  (\ref{r14a}) since $\rho^{-1}=d I_d$.

It is interesting to compare the maximum confidence $C$ with the confidence
$C_{ME}$ that is achieved in minimum-error discrimination, where inconclusive results do not occur and the probability of correct results, $P_{corr}$, is maximal.
For symmetric pure states the
minimum-error measurement is known to be the square-root measurement
described by the detection operators  \cite{ban}
\begin{equation}
\label {r15} \Pi_{j}^{ME}=
\frac{1}{N}\rho^{-1/2} |\psi_j\rangle\langle \psi_j |\rho^{-1/2}=
\frac{d}{N} V^{j} |\nu_1\rangle\langle \nu_1|V^{\dag j}
\end{equation}
where for the second
equality sign we applied Eqs. (\ref{r10a}) and (\ref{r13}).  
 From  Eq. (\ref{s5a}) with $Q=0$ we obtain $P_{corr}^{ME}=C_{ME}$. Using  the 
operators  $ \Pi_{j}^{ME}$ and     
 $\eta_j=1/N$ we arrive at  
\begin{equation}
\label {r16} C_{ME} = \frac{d}{N}|\langle \psi_1|\nu_1\rangle|^2=C |\langle \psi_1|\nu_1\rangle|^2=
\frac{C}{d}\left(\sum_{l=1}^d|c_l|\right)^2
\end{equation}
which shows that  $C_{ME}\leq C$ as expected.
 Obviously the  gain in confidence achieved  by admitting
inconclusive results and performing a maximum-confidence measurement
depends on the expansion coefficients $c_l$ of the given pure states with
respect to the eigenbasis of the symmetry operator.
Equality only holds when $|c_l|=1/\sqrt{d}$ for each value of $l$.
In this  case Eq. (\ref{r14}) shows that there are no inconclusive
results in the maximum-confidence measurement. In addition,  it
follows that the detection operators in Eqs. (\ref{r14a}) and
(\ref{r15}) are identical, that is the minimum-error measurement and the optimized maximum-confidence measurement 
coincide.

\subsubsection {$N$ symmetric mixed  qubit states}

In our second example we consider the discrimination of $N$ symmetric mixed states of rank
2 in a two-dimensional joint Hilbert space. Since an arbitrary
mixed qubit state can be always written in the form of  Eq.
(\ref{r10}) with $d =2$ and with a certain value of the parameter
$p$ $(p\neq 0)$, the maximum confidence $C$
corresponds to the largest eigenvalue
 of the operator $\tilde{\rho}_1$ in  Eq. (\ref{r12}) with $d=2$.
  Upon
 determining the spectral representation of this operator, using
the expansion  $|\psi_1\rangle= c_1|r_1\rangle + c_2|r_2\rangle$, where $|r_1\rangle$  
and  $|r_2\rangle$ are the eigenstates of the symmetry operator $V,$ we obtain
 both the largest eigenvalue, equal to  the maximum confidence,  
\begin{equation}
\label {r18}  C= \frac{1}{N} \left[1+\frac{p\,|c_1
c_2|}{\sqrt{\left(p|c_1|^2+\frac{1-p}{2}\right)\left(p|c_2|^2+\frac{1-p}{2}\right)}}\right]
\end{equation}
and the  corresponding eigenstate  $|\nu_1\rangle =
\frac{1}{\sqrt{2}}(|r_1\rangle + |r_2\rangle)$. Equation
 (\ref{r5a}) yields
\begin{equation}
\label {r19} \alpha_{opt}= \frac{1}{N}\left(1-p+2p\,{\rm
Min}\{|c_1|^2\!,|c_2|^2 \}\right)
\end{equation}
which in turn determines  the optimum detection operator
$\Pi_1$, see Eq. (\ref{r4}),  as well as  the minimum
failure probability necessary for maximum-confidence discrimination, 
\begin{eqnarray}
\label {r20}  Q_{min} = p\left(1- 2\,{\rm min}\{|c_1|^2\!,|c_2|^2
\}\right).
\end{eqnarray}
Introducing  
\begin{equation}
\label {r20a}
|c_1|= \cos\frac{\gamma}{2}, \quad |c_2|=
\sin\frac{\gamma}{2}\quad {\rm with}\;\, 0\leq \gamma \leq \pi/2,
\end{equation}
 that is, 
${\rm min}\{|c_1|^2\!,|c_2|^2
\} =\sin^2 {\frac{\gamma}{2}}$, the above equations
can be rewritten as
\begin{equation}
\label {mix-sym3} C= \frac{1}{N}\left(1
+\frac{p\sin\gamma}{\sqrt{1-p^2\cos^2\gamma}}\right),\quad Q_{min} =
p\cos \gamma.
\end{equation}
Clearly, the maximum value of the confidence decreases
 with growing number of
states while the minimum failure probability necessary for maximum-confidence discrimination 
stays constant. 

We note that two arbitrary mixed
qubit states in the same Hilbert space and with the same purity always belong to the class of symmetric states. 
They can be represented by using  
Eqs. (\ref{r10}) and  (\ref{r10a}) with $N=d=2$ and with $|\psi_{1,2}\rangle =
\cos^2\frac{\gamma}{2}|0\rangle \pm e^{i\phi}\sin^2
\frac{\gamma}{2}|1\rangle$  where the states  $|0\rangle$ and $|1\rangle$ are orthonormal basis states   
and the symmetry operator is given as $V=|0\rangle \langle 0| - |1\rangle \langle 1|$.
With       
$ |\langle \psi_1|\psi_2\rangle|= \cos \gamma $    
  Eq. (\ref{mix-sym3}) corresponds to  our earlier result for the
maximum confidence discrimination of two equally probable mixed
qubit states having the same purity \cite{herzog1}.

\subsubsection {$N$ special symmetric mixed  states in a $d$-dimensional joint Hilbert space $(N\geq d)$}

Our last example refers to  the case that  in Eq. \eqref{r10} $p\neq 0$ and the dimension $d$ of the
joint Hilbert space is arbitrary. However, we assume that  the
$N$ states $|\psi_j\rangle$ are of the special kind where
the modulus of all expansion coefficients is the same, that is, where Eqs. (\ref{r11a}) -- (\ref{r12}) take the form 
\begin{equation}
\label {special1} |\psi_1\rangle=
\sum_{l=0}^d \frac{|r_l\rangle}{\sqrt{d}},\quad \rho= \frac{I_d}{d},\quad
\tilde{\rho}_1=\frac{I_d(1-p)}{N} +\frac{p d}{N}
|\psi_1\rangle\langle \psi_1|.
\end{equation}
The largest eigenvalue of $\tilde{\rho}_1$,  belonging to the eigenstate $|\nu_1\rangle=
|\psi_1\rangle$ and determining the maximum confidence, $C$,
 can be immediately
read out. Using  Eqs. (\ref{r4}) -- (\ref{r5a})  we find that
\begin{equation}
\label {special2} C= \frac{1+p(d-1)}{N}, \quad
\alpha_{opt}=\frac{1}{N},\quad Q_{min}=0.
\end{equation}
 The optimum detection operators are
\begin{eqnarray}
\label {special3} 
\Pi_j 
=\frac{d}{N}
|\psi_j\rangle\langle \psi_j|\quad (j=1,\ldots,N)
\end{eqnarray}
where Eq. (\ref{r14a}) with $\rho=I_d/d$ has been taken into account. Clearly, in this special case the optimized maximum-confidence measurement for discriminating the given
 mixed states
does not require  inconclusive results, that is, $\Pi_0=0.$ Comparison with Eq. (\ref{r15}) shows
that the measurement    is equal to the
minimum-error measurement for discriminating the underlying pure states
$|\psi_j\rangle$.

\section{Summary and concluding remarks}

To summarize, in this paper we derived  necessary and sufficient
optimality conditions for a measurement which discriminates $N$
mixed quantum states with maximum confidence for each conclusive
outcome, thereby keeping the overall probability of inconclusive
outcomes as small as possible. These conditions are given by  Eqs.
(\ref{Z1}) and (\ref{Z3})
 together with Eqs. (\ref{requ}) and (\ref{Lambda-j-a}). They generalize earlier optimality conditions \cite{eldar} which refer to the
 special case of optimum unambiguous discrimination. 
 We derived general properties of the optimum measurement and applied the optimality conditions to  equiprobable symmetric states. 
 For these states we presented analytical solutions of the optimization problem for examples where the detection operators describing the  maximum-confidence discrimination are one-dimensional.

When higher-rank detection operators are involved, the general problem of minimizing the failure probability gets similarly complicated for maximum-confidence discrimination  as it is for unambiguous discrimination. As shown in Sec.III E, when  $N=2$ both problems are mathematically equivalent. It has been found that already for the simplest general case of  higher-rank detection operators in unambiguous discrimination,  that is for the  discrimination of two density operators of rank 2 in a  four-dimensional Hilbert space, the optimization problem can in general lead to polynomial equations of higher degree \cite{kleinmann}. 
 For $N\geq 3$ a general solution is not known  for the unambiguous discrimination of mixed states and only bounds have been derived \cite{feng1,zhang}. 
However,  analytical solutions with higher-rank detection operators can be
 constructed in special cases where the given density operators
allow us to separate the optimization problem into independent
 optimization problems  in mutually orthogonal subspaces
of the Hilbert space and where the projections of the detection operators onto these subspaces are one-dimensional.  This method has been applied for the optimum unambiguous discrimination with $N=2$,  see, e. g., \cite {HB,BFH,herzog}, and also for a case of   equiprobable symmetric mixed states with arbitrary $N$ \cite{HBqudit},  as well as  for the optimized maximum-confidence discrimination of two mixed states, where an example was given in  \cite{herzog1}. 

We still remark that other state-discrimination strategies have been introduced where the overall probability of getting a correct result,  $P_{corr}$, is maximized under the constraint that either the failure probability  \cite{chefles-barnett,fiurasek,eldar1} or the error probability 
\cite{touzel,hayashi-err,sugimoto} has a certain fixed value. Maximum-confidence discrimination is related to the former of these,  
  as discussed already in  \cite{herzog1} and \cite{barnett}.  In fact, when for the given states the maximum achievable confidence is the same for each individual outcome,  that is, when    $C_j=C$ for $j=1,\ldots,N$,  the maximum-confidence measurement where the failure probability takes its minimum,  $Q_{min}$,   coincides with the measurement which  maximizes  $P_{corr}$  under the condition that  $Q$ is fixed at the value $Q_{min}$. This is due to the fact that the latter measurement also maximizes the ratio $P_{corr}/{(1-Q)}$ at the fixed value of $Q$ and that according to Eq. (\ref{s5a}) this ratio is equal to the confidence.  Hence  in this case  maximizing $P_{corr}$ at a fixed value $Q$ with  $0 \leq Q\leq Q_{min}$ 
 corresponds to interpolating between minimum-error discrimination for  $Q=0$, and optimized maximum-confidence discrimination for $Q=Q_{min}$, or optimum unambiguous discrimination, respectively, if  $C=1$.

On the other hand, when the maximum confidence differs for the individual outcomes,  it 
follows that the maximum of $P_{corr}/(1-Q)$ at a fixed value of $Q$ is equal to  ${\rm max}_j \{C_j\}$,  and that it is  obtained in a modified maximum-confidence measurement where all states $j$ with  $C_j< {\rm max}_j \{C_j\} $ yield an inconclusive result \cite{herzog1}. The  failure probability resulting from a  measurement of this kind is not necessarily the smallest one that can be reached in maximum-confidence discrimination.  For two mixed qubit states with arbitrary values of $C_1$ and $C_2$, defined in the same Hilbert space,  maximum-confidence discrimination with minimum failure probability has been studied in our earlier  paper \cite{herzog1}.    
 In order to obtain solutions for discriminating more than two states,  the optimality conditions derived in this paper can be applied.

\begin{acknowledgments}
Discussions on various aspects of state discrimination with Janos
Bergou, Oliver Benson and Gesine Steudle are gratefully
acknowledged.
\end{acknowledgments}

\appendix*
\section{}

We want to show that Eq. (\ref{Z1}) is not only sufficient, but also necessary for Eq.
(\ref{dual2}) to hold.  In analogy to a recent treatment of
minimum-error discrimination  \cite{barnett-croke}, we perform the
proof  by demonstrating an example where a single negative eigenvalue of one of
the operators in Eq. (\ref{Z1}) leads to a violation of Eq.
(\ref{dual2}).

Let us first assume that a particular one of the operators
$\Lambda_j(Z-\rho)\Lambda_j$, say the one with $j=N$, has a negative
eigenvalue $-\mu$, resulting in the eigenvalue equation
\begin{equation}
\label {app1}   \Lambda_N(Z-\rho)\Lambda_N |\mu\rangle = -\mu
|\mu\rangle\quad {\rm with} \;\,\mu > 0.
\end{equation}
Now we suppose that the detection  operators $\Pi_0,\ldots, \Pi_N$ are
optimal which means that they  obey   the equalities  in Eq.
(\ref{Z3}). In analogy to \cite{barnett-croke} we define another set of operators,
given as
\begin{equation}
\label {app2} \Pi_j^{\prime}=
\left(I_d-\epsilon|\mu\rangle\langle\mu|\right)
 \Pi_j (I_d-\epsilon|\mu\rangle\langle\mu|)
\end{equation}
for $j=0,1,\ldots,N-1$ and
\begin{equation}
\label {app3} \Pi_N^{\prime}=  (I_d-\epsilon|\mu\rangle\langle\mu|)
\Pi_N (I_d-\epsilon|\mu\rangle\langle\mu|) + \epsilon(2-
\epsilon)|\mu\rangle\langle\mu|
\end{equation}
where $0\leq \epsilon \ll 1$.  It can be easily checked by a straight-forward calculation that the
primed operators fulfill the conditions for completeness and
positivity expressed by Eq. (\ref{cond1}) and are thus a valid set
of detection operators, yielding the discrimination
 probability
$R^{\prime} = \sum_{j=1}^N {\rm Tr}(\rho
\Lambda_j\Pi_j^{\prime}\Lambda_j)$. Using Eqs. (\ref{app1}) --
(\ref{app3}) as well as the completeness relation $\sum_{j=0}^N \Pi_j^{\prime}=I_d$ we obtain
\begin{eqnarray}
\label {app4} \lefteqn{{\rm Tr} Z - R^{\prime}= {\rm Tr
}(Z\Pi_0^{\prime})
+ \sum_{j=1}^N {\rm Tr }[\Lambda_j(Z-\rho)\Lambda_j\Pi_j^{\prime}]}\\
&&= \epsilon(2- \epsilon){\rm Tr}[\Lambda_N
(Z-\rho)\Lambda_N|\mu\rangle\langle\mu|]= - 2\epsilon\mu + O(
\epsilon^2)\nonumber
\end{eqnarray}
 where for the second equality sign  Eq. (\ref{Z3})
has been taken into account. Hence the negativity of
$\Lambda_N(Z-\rho)\Lambda_N$
   implies that ${\rm Tr} Z - R^{\prime}
<0$, in contrast to Eq. (\ref{dual2}). Thus we have shown that the
positivity of all operators $\Lambda_j(Z-\rho)\Lambda_j$ is a
necessary condition for  Eq. (\ref{dual2}).

With respect to the positivity of $Z$ the proof proceeds in a
completely analogous way. We now assume that
\begin{equation}
Z|\mu\rangle = -\mu
|\mu\rangle \quad {\rm
 with}\;\;\mu > 0
\end{equation}
 and we suppose that the primed detection
operators are determined by Eq. (\ref{app2}) for $j=1,\ldots,N$
while $\Pi_0^{\prime}$ is given by Eq. (\ref{app3}) with $N$
replaced by 0. Applying   Eq. (\ref{Z3}) we then again find that
\begin{equation}
{\rm Tr} Z - R^{\prime} =  - 2\epsilon\mu + O( \epsilon^2) <0,
\end{equation}
which means that the negativity of $Z$ leads to a violation of Eq.
(\ref{dual2}), or, in other words,  that the positivity of $Z$ is a
necessary condition for Eq. (\ref{dual2}).


\begin{thebibliography}{99}

\bibitem{helstrom}
    C. W. Helstrom, {\it Quantum Detection and Estimation Theory}
    (Academic, New York, 1976).

\bibitem{ivan} I. D. Ivanovic, Phys. Lett. A {\bf 123}, 257
  (1987).

\bibitem{dieks} D. Dieks, Phys. Lett. A {\bf 126}, 303 (1988).

\bibitem{peres} A. Peres, Phys. Lett. A {\bf 128}, 19 (1988).

\bibitem{jaeger} G.\ Jaeger and A.\ Shimony, Phys.\ Lett.\ A {\bf 197},
  83 (1995).

\bibitem{rudolph} T. Rudolph, R. W. Spekkens, and P. S. Turner,
  Phys. Rev. A {\bf 68}, 010301(R) (2003).

\bibitem{raynal} Ph. Raynal, N. L\"utkenhaus, and S. van Enk,
  Phys. Rev. A {\bf 68}, 022308 (2003).

\bibitem{eldar}Y. C. Eldar, M. Stojnic, and B. Hassibi, \pra {\bf 69},
062318 (2004).

\bibitem{HB} U. Herzog and J. A. Bergou, \pra {\bf 71},
050301(R) (2005).

\bibitem{BFH}
J. A Bergou, E. Feldman, and M. Hillery, \pra {\bf 73}, 032107
(2006).

\bibitem{herzog} U. Herzog, \pra {\bf 75}, 052309 (2007).

\bibitem{raynal2} Ph. Raynal and N. L\"utkenhaus, \pra {\bf 76}, 052322 (2007).

\bibitem{kleinmann} M. Kleinmann, H. Kampermann, and D. Bru\ss,
Phys. Rev. A {\bf 81}, 020304(R) (2010); J. Math. Phys. {\bf 51}, 032201 (2010). 

\bibitem{feng1} Y. Feng, R. Duan, and M. Ying, \pra {\bf 70}, 012308 (2004).

\bibitem{zhang} C. Zhang, Y. Feng, M. Ying,  Phys. Lett. A {\bf 353}, 300 (2006).

\bibitem{chefles1} A. Chefles, Phys. Lett. A {\bf 239}, 339 (1998).

\bibitem{support}
The support of a density operator is the Hilbert space spanned by
its eigenvectors with nonzero eigenvalues. The rank is the dimension of the support. 

\bibitem{croke} S. Croke, E. Andersson, S. M. Barnett, C. R. Gilson
and J. Jeffers, Phys. Rev. Lett. {\bf 96}, 070401 (2006).

\bibitem{croke1} S. Croke, E. Andersson and S. M. Barnett, \pra {\bf 77}, 012113 (2008).

\bibitem{mosley} P. J. Mosley, S. Croke, I. A. Walmsley and S. M.
Barnett, \prl {\bf 97}, 193601 (2006).

\bibitem{herzog1} U. Herzog, Phys. Rev. A {\bf 79}, 032323 (2009).

\bibitem{herzog-benson} U. Herzog and O. Benson,  J. Mod. Opt {\bf 57}, 188
(2010).

\bibitem{steudle} G. A. Steudle, S. Knauer, U. Herzog, E. Stock, V.
Haisler, D. Bimberg, and O. Benson, Phys. Rev. A {\bf 83}, 050304(R) (2011).

\bibitem{jimenez}  O. Jim\'enez, M. A. Solis-Prosser, A. Delgado, and L. Neves,
Phys. Rev. A {\bf 84}, 062315  (2011).

\bibitem{barnett-croke} S. M. Barnett and S. Croke, J. Phys. A
 {\bf 42}, 062001 (2009).

\bibitem{ban}  M. Ban, K. Kurokawa, R. Momose, and O. Hirota, Int. J.
Theor. Phys. {\bf 36}, 1269 (1997).

\bibitem {chefles-symm} A. Chefles and S. M. Barnett, Phys. Lett. A {\bf 250}, 223
(1998).

\bibitem{HBqudit}
U. Herzog and J.  A.  Bergou,  Phys. Rev. A {\bf 78}, 032320 
(2008), Erratum: Phys. Rev. A {\bf 78}, 069902(E) (2008).

\bibitem{chefles-barnett}  A. Chefles and S. M. Barnett,
J. Mod. Opt. {\bf 45}, 1295 (1998).

\bibitem{fiurasek}
J. Fiur\'a\v{s}ek and M. Je\v{z}ek, \pra {\bf 67}, 012321 (2003).

 \bibitem{eldar1}
 Y. C. Eldar, \pra {\bf 67}, 042309 (2003).

\bibitem{touzel}
M. A. P. Touzel, R. B. A. Adamson, and A. M. Steinberg, \pra {\bf
76}, 062314 (2007).

\bibitem{hayashi-err}
A. Hayashi,  T. Hashimoto, and M. Horibe,  \pra {\bf 78},
012333 (2008).


\bibitem{sugimoto}
H. Sugimoto, T. Hashimoto, M. Horibe, and A. Hayashi, \pra {\bf 80},
052322 (2009).

\bibitem{barnett}
S. M. Barnett and S. Croke, Adv. Opt. Photon. {\bf 1}, 238 (2009).

\end{thebibliography}
\end{document}